\documentclass[aps,prl,twocolumn,superscriptaddress,amsmath,amsfonts,showpacs]{revtex4}

\usepackage{graphicx}

\begin{document}


\title{Fractional flux quanta at intrinsic metallic interfaces of \\
noncentrosymmetric superconductors}


\author{C. Iniotakis}
\affiliation{Institute for Theoretical Physics,  ETH Zurich, 8093 Zurich, Switzerland}
\author{S. Fujimoto}
\affiliation{Department of Physics, Kyoto University, Kyoto 606-8502, Japan}
\author{M. Sigrist}
\affiliation{Institute for Theoretical Physics, ETH Zurich, 8093 Zurich, Switzerland}




\date{\today}

\begin{abstract}
We examine intrinsic  interfaces separating crystalline 
twin domains of opposite spin-orbit coupling
in a noncentrosymmetric superconductor such as CePt$_3$Si. At these interfaces,
low-energy Andreev bound states occur
as a consequence of parity-mixed Cooper pairing, and
a superconducting phase which violates time reversal symmetry can be
realized. This provides an environment allowing flux lines with fractional
flux quanta to be formed at the interface. Their presence could have strong 
implications on the flux creep behavior in such superconductors. 
 \end{abstract}



\maketitle



Symmetry is a decisive factor for many properties of materials.
Lowering a symmetry can yield new couplings between physical 
observables and causes intriguing phenomena.
The recently discovered noncentrosymmetric superconductors CePt$_3$Si,
CeRhSi$_3$, CeIrSi$_3$, and Li$_2$(Pt$_x$Pd$_{1-x}$)$_3$B provide such
examples \cite{Bauer,Kimura,Sugitani,Togano}. In these materials,
the absence of an inversion center generates antisymmetric spin-orbit 
interaction and leads, in the superconducting state, to 
parity-mixing of Cooper pairs, magnetoelectric effects, 
and many other interesting features  \cite{Edelstein,Gorkov,Frigeri,Fujimoto}.
In many cases, such crystal structures permit the existence of twin domains
 exhibiting opposite inversion symmetry breaking within a single crystal. 
Actually, in the crystal growth processes of noncentrosymmetric materials,
the formation of such twin domains is inevitable.
The existence of twin domains in noncentrosymmetric superconductors is also
suggested by a recent experiment, which revealed
that a high quality single crystal sample of CePt$_3$Si exhibits a lower transition temperature than 
polycrystal ones \cite{Takeuchi}. 
Since the origin of this behavior cannot be understood in terms of 
conventional impurity effects  \cite{Mineev}, possibly 
twin boundaries could enhance the trend to superconductivity. 
Furthermore, recent NMR measurements of the single crystal sample are
 ingeniously interpreted by assuming the existence of  
twin domains \cite{Mukuda}. Motivated by these observations, in this letter, we investigate
effects of intrinsic interfaces between twin domains
on the parity-mixed superconducting state. 
Our central finding is, that superconducting states with broken time-reversal 
symmetry can occur at the interfaces, allowing for fractional vortices. 

We consider a noncentrosymmetric superconductor such 
as CePt$_3$Si and assume for simplicity a spherical Fermi surface 
parametrized by the unit vector $\hat{\bf{k}}=(\cos\varphi \sin\theta, \sin\varphi \sin\theta,\cos\theta )$.
The presence of a Rashba-type spin-orbit coupling, $\alpha (\hat{\bf z} \times \hat{ \bf{k}}) \cdot \bf{s} $,
induces a splitting of the electron bands and the Fermi surface into sheets,
each exhibiting a specific spin structure. The superconducting phase displays a mixed parity 
\cite{Edelstein,Gorkov,Frigeri}, and the state compatible with experiments consists of an
$s$- and a $p$-wave component, being of $ s \pm p $-character on the two Fermi sheets. 
Moreover, there is experimental evidence for a nodal gap structure, which suggests a dominant spin-triplet $p$-wave 
component with $ q= \Delta_s/\Delta_p < 1 $,
where $ \Delta_s $ and $ \Delta_p $ denote the magnitudes of the $s$- and the $ p$-wave components
in the superconducting gap \cite{Bonalde,Hayashi}. 
For the calculations, we employ quasiclassical Eilenberger theory
of superconductivity \cite{Eilenberger,Larkin,Serene}. 
This method provides a convenient and powerful tool 
for describing superconductivity and
has been applied to noncentrosymmetric superconductors in Ref. \cite{Hayashi}.  
According to Ref. \cite{Iniotakis}, the superconducting state can be expressed by 
the so-called bulk coherence functions $\gamma_B,\tilde\gamma_B$ straightforwardly, 
which corresponds to the Riccati formulation of Eilenberger theory \cite{Schopohl, Eschrig}. 
Using an effective one-band description,  where the size of the band splitting is assumed to 
be small compared to the Fermi energy, we obtain \cite{Iniotakis}
\begin{subequations}
\label{EQBulkGamma}
\begin{eqnarray}
\gamma_B&=&-(\gamma_+ \hat\sigma_+ + \gamma_- \hat\sigma_-)\hat \sigma_y \\
\tilde \gamma_B&=&\hat\sigma_y (\gamma_+ \hat\sigma_+ + \gamma_- \hat \sigma_-),
\end{eqnarray}
\end{subequations}
where the coefficients are defined as
\begin{equation}
\label{EQgammaPM}
\gamma_\pm=\frac{\Delta_\pm}{\omega_n +\sqrt{\omega^2_n+|\Delta_\pm|^2}}
\end{equation}
with $ \Delta_\pm (\hat{\bf{k}}) =\Delta_s \pm \Delta_p \sin \theta $ 
and $\omega_n=(2n+1)\pi k_B T$ denoting Matsubara frequencies. While we neglect the splitting
of the bands,  we keep their spin structure imposed by the spin-orbit coupling, as described by
\begin{equation}
\hat{\sigma}_\pm=\frac{1}{2}
\left( 
\begin{array} {cc} 1 & \mp i e^{-i \varphi} \\ \pm i e^{i \varphi}  & 1 \end{array}
\right).
\end{equation}
These spin matrices have the useful projection properties $ \hat\sigma_\pm^2 = \hat\sigma_\pm $ , 
$ \hat\sigma_+ \hat\sigma_-= \hat\sigma_- \hat\sigma_+=0 $ and $ \hat\sigma_+ +\hat\sigma_- = 1 $. 

\begin{figure}[t]
\includegraphics[width=0.95 \columnwidth]{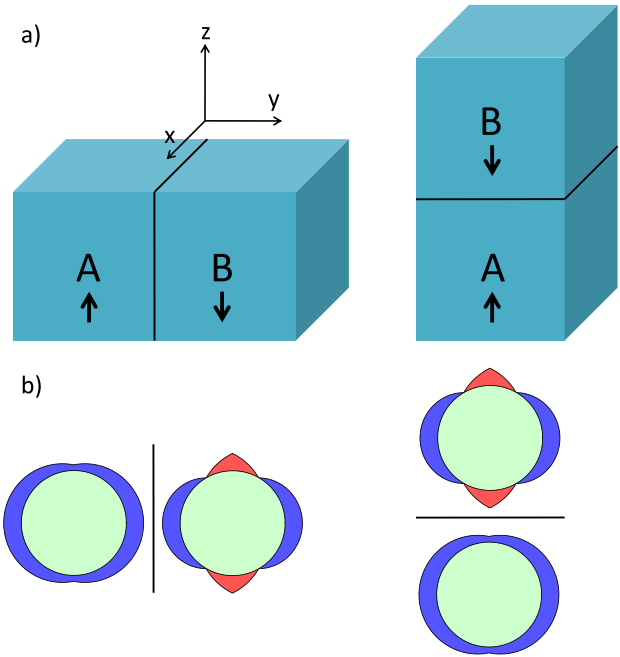}
\caption{\label{Fig01}
Sketch of two intrinsic metallic interfaces in a noncentrosymmetric
superconductor with normal vector $\hat{\mathbf{n}}$ perpendicular (left) or
parallel (right) to the $z$-axis. a) The interface separates regions $A$ and $B$, which 
exhibit a different direction of the spin-orbit coupling  $\alpha$
as indicated by the arrows. b) The situation is analogous to a junction between two ordinary 
singlet superconductors with gap amplitudes $\Delta_+$ and $-\Delta_-$. The sign of the
gap amplitude is illustrated by the two different colors. 
}
\end{figure}

We now turn to the electronic properties of an interface separating  regions $A$ and $B$, which
are characterized by the opposite sign of the antisymmetric spin orbit coupling according to
$ \alpha_A = - \alpha_B $. 
An illustration of two specific situations for such an interface 
can be found in Fig. \ref{Fig01}, a), and similar setups have already been examined in
a different context \cite{Borkje1,Borkje2}. For our following analysis we neglect the direct influence of the
interface on the superconducting order parameter and fix the moduli of the pair potentials to remain constant
even at the interface.
This approximation does not affect our discussion qualitatively, and quantitative corrections
are minor. It leads to the simplification, however, that  the coherence functions at the interface 
can be replaced by the corresponding bulk coherence functions. 
In the following, we use   bulk coherence functions $\gamma^A$ and $\tilde\gamma^A$ for region $A$
according to Eqs. (\ref{EQBulkGamma}). Regarding region $B$, one might be tempted to
get the bulk coherence functions $\gamma^B$ and $\tilde\gamma^B$ by simply
interchanging the two gap amplitudes $\Delta_\pm$. For small values of $q$ it is
rather natural, however, to keep the dominant $p$-wave component $\Delta_p$ constant on both
sides of the interface. Then,  the $s$-wave component changes its sign across the
interface and an additional phase factor $-1$ has to be  introduced on side $B$.
We allow the gap function of region $B$ to exhibit a further general phase
difference with respect to region $A$, which is denoted by $\phi$  in the following.
The bulk coherence functions of region $B$ are then given by
\begin{subequations}
\begin{eqnarray}
\gamma^B&=&(\gamma_- \hat\sigma_+ + \gamma_+ \hat\sigma_-)\hat \sigma_y e^{i \phi}\\
\tilde \gamma^B&=&-\hat\sigma_y (\gamma_- \hat\sigma_+ + \gamma_+ \hat \sigma_-) e^{-i \phi}.
\end{eqnarray}
\end{subequations}

Generally, once  the coherence functions $\gamma,\tilde \gamma$ are known  for
a specific Fermi vector at a given point in space, also the quasiclassical 
Green's function $\hat g$ in $2\times 2$ spin space is immediately available as
\begin{equation}
\hat g= (1-\gamma \tilde\gamma)^{-1} (1+\gamma \tilde\gamma) = 2(1-\gamma \tilde\gamma)^{-1}-1.
\end{equation}
The interface is implemented by  well-established boundary conditions for the Green's function or the coherence
functions, respectively \cite{Zaitsev, Shelankov, Eschrig}. Restricting ourselves to
a high-transparency interface, the resulting  Green's function directly at the interface is
given by
\begin{eqnarray}
\nonumber
\hat g&=&2(1-\gamma^A \tilde \gamma^B)^{-1}-1\\
\nonumber
&=&2[1-(\gamma_+ \hat\sigma_+ + \gamma_- \hat\sigma_-) (\gamma_- \hat\sigma_+ + \gamma_+ \hat \sigma_-) e^{-i \phi}]^{-1}-1\\
\nonumber
&=&2[1-(\gamma_+ \gamma_- \hat\sigma_+ + \gamma_- \gamma_+ \hat\sigma_-) e^{-i \phi}]^{-1}-1 \\
\label{EQgInterface}
&=&\frac{2}{1-\gamma_+ \gamma_-  e^{-i \phi}}-1,
\end{eqnarray}
where the projection properties of $\hat\sigma_\pm$ have been used in the
intermediate steps. Note, that the expression  Eq. (\ref{EQgInterface}) for the quasiclassical
Green's function at the interface only holds for quasiparticle trajectories
with Fermi vectors $\hat {\bf k}$ pointing from $A$ to $B$. In the opposite case,
the superscripts $A$ and $B$ have to be interchanged, and we find the symmetry
relation $\hat g(\omega_n,-\hat{\mathbf{k}} )=\hat g(\omega_n,\hat{\mathbf{k}} )^*$.
Several points should be mentioned here.
Firstly, only bands having the same spin structure, $\hat\sigma_+$ or $\hat\sigma_-$, contribute to this Green's function, 
the other combinations vanish by projection.  Furthermore, the Green's function is proportional to
the unit matrix. As a consequence, there is an analogy between  this interface of noncentrosymmetric
superconductors and a standard interface consisting of  two singlet superconductors 
as illustrated in Fig. \ref{Fig01}, b). 

In the following, we derive the Josephson current density through the interface.
Using the symmetry relation stated above, we find
\begin{equation}
\mathbf{j} (\phi)=4\pi e N_0 k_B T v_F \sum_{\omega_n >0}^{\omega_c} \langle \hat{\bf k} \text{Im}[g] \rangle,
\end{equation}
where $\langle...\rangle$ denotes averaging over half of the Fermi sphere determined by quasiparticle directions 
$\hat {\mathbf{k}}$  pointing from region $A$ to $B$, 
and $g$ is the unit matrix component of $\hat g$ according to Eq. (\ref{EQgInterface}). Written in normalized
quantities $\hat{T}=T/T_c$ and $\hat j=j/4\pi e N_0 k_B T_c v_F$, we eventually find the result
\begin{equation}
\label{EQCurrent}
\hat{\bf j}(\phi) 
=\hat{T}\sum_{\omega_n >0}^{\omega_c} \int_0^{\frac{\pi}{2}} d\theta 
\left\{
\begin{array}{r}    \frac{4}{\pi} \sin^2\theta  \\  \sin 2 \theta               \end{array}  
\right\} 
\text{Im}\left[\frac{1}{1-\gamma_+ \gamma_-  e^{-i \phi}}\right] \hat{ \mathbf{n}},
\end{equation}
where the upper (lower) formula stands for the situation with the normal vector $\hat{\bf{n}}$ of the interface 
perpendicular (parallel) to the $z$-axis of the system. In both cases  only components of the  current flowing perpendicular
to the interface are allowed by symmetry.  The values of $\gamma_+, \gamma_-$ in the integrand
are real and depend on $\sin\theta$ themselves [cf. Eq. (\ref{EQgammaPM})].

\begin{figure}[t]
\includegraphics[width=0.95 \columnwidth]{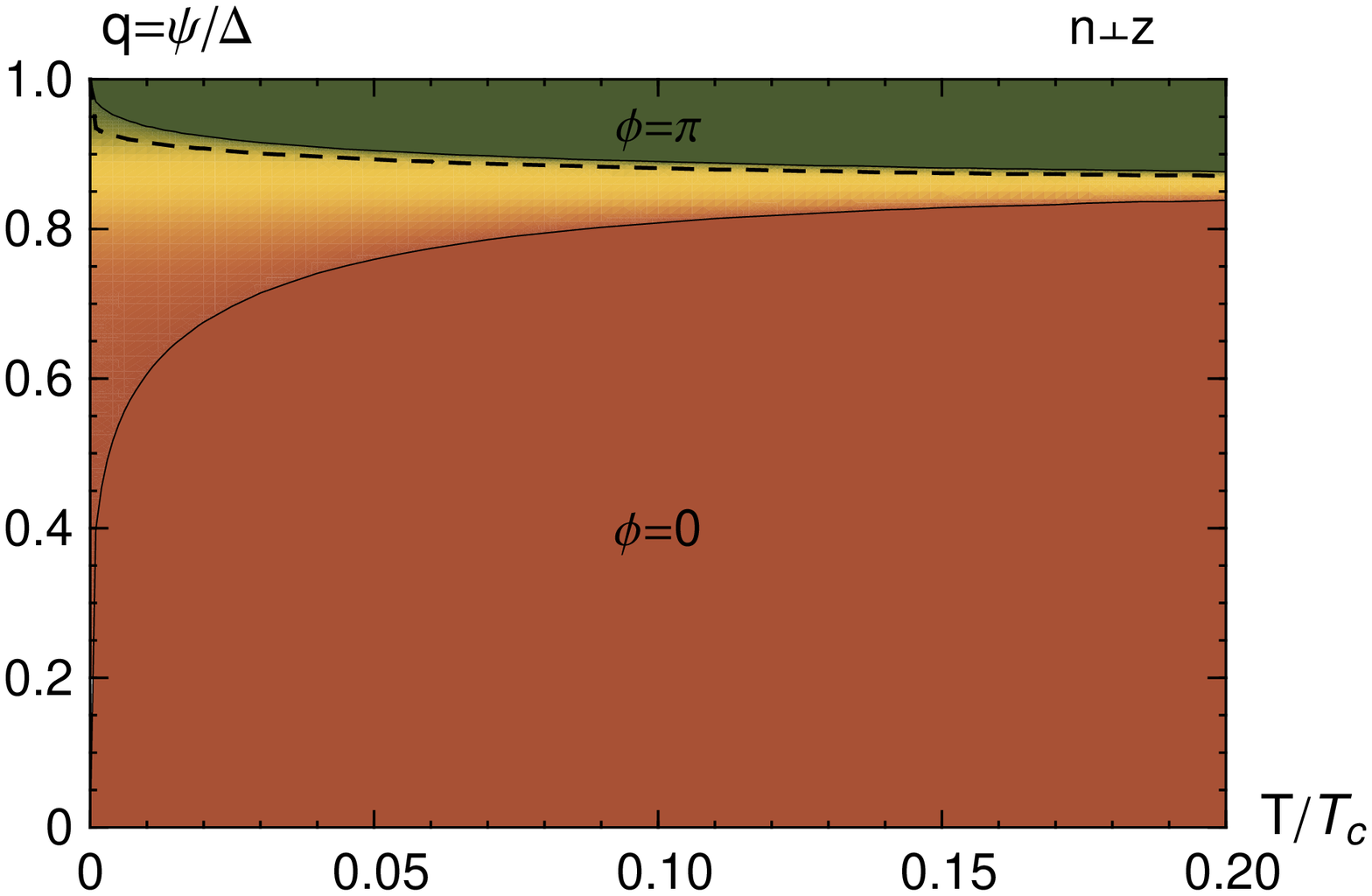}
\caption{\label{Fig02}
(Color online) Part of the $q,T$-phase diagram of an intrinsic metallic junction in a 
noncentrosymmetric superconductor. There are 
regions of a stable Josephson phase $\phi=0$ (bottom) and $\phi=\pi$ (top).
In between, a phase difference $0<\phi<\pi$ is favorable.
Here, the normal vector $\hat{\mathbf{n}}$ of the interface is taken to be
perpendicular to the $z$ axis, corresponding to the lefthand scenario of Fig. \ref{Fig01}.
For $T\rightarrow T_c$, $0$- and $\pi$-phase meet at $q=\sqrt3/2\approx0.87$.
The dashed line indicates the boundary between $0$- and $\pi$-phase in
the contrary scenario of a  low-transparency tunnel junction.
}
\end{figure}

\begin{figure}[t]
\includegraphics[width=0.95 \columnwidth]{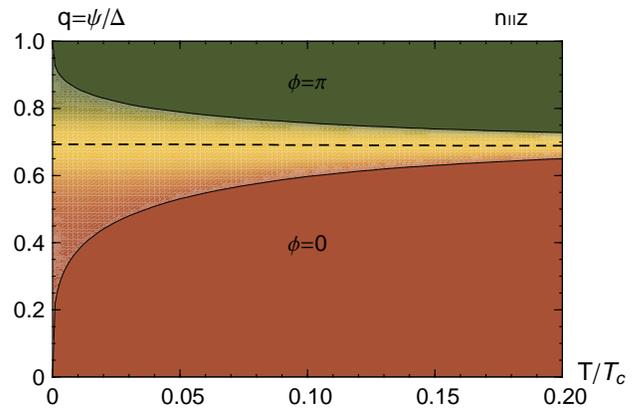}
\caption{\label{Fig03}
(Color online) Part of the $q,T$-phase diagram of an intrinsic metallic junction in a 
noncentrosymmetric superconductor. There are 
regions of a stable Josephson phase $\phi=0$ (bottom) and $\phi=\pi$ (top).
In between, a phase difference $0<\phi<\pi$ is favorable.
Here, the normal vector $\hat{\mathbf{n}}$ of the interface is taken to be
parallel to the $z$ axis, corresponding to the righthand scenario of Fig. \ref{Fig01}.
For $T\rightarrow T_c$, $0$- and $\pi$-phase meet at $q=1/\sqrt 2\approx 0.71$.
The dashed line indicates the boundary between $0$- and $\pi$-phase in
the contrary scenario of a low-transparency tunnel junction.
}
\end{figure}

Numerical evaluation of the current-phase relations according to Eq. (\ref{EQCurrent})
allows us to determine the phase difference $\phi$ of the stable interface states. 
We focus on $\phi>0$, keeping in mind that with $\phi$ also $-\phi$ is a stable solution.
For the two situations depicted in Fig. \ref{Fig01} we can derive the phase diagrams 
displayed in Figs. \ref{Fig02} and \ref{Fig03}, respectively. 
We find three regions in the $ q$-$T$-phase diagram. In the region of small $ q $, the stable
state corresponds to $ \phi =0 $ and for $ q $ close to one it is $ \phi =  \pi $.
 The latter means,  that the $s$-wave component $ \Delta_s $ would remain
unchanged across the interface. 
 Intriguing is a region in between these two limits, where the stable phase difference has 
an intermediate value $ 0 < \phi < \pi $. 
Note, that the $ \pm \phi $ solutions are degenerate for this intermediate region, reflecting the  
fact that such an interface state is time-reversal symmetry breaking, since $\phi$ changes 
sign under the time reversal operation.  
The intermediate region of $ q $-values shrinks with increasing temperature,
eventually reaching a single point at $ T_c $. For the normal vector perpendicular (parallel) to $z$ this 
 value is $q=\sqrt3/2\approx0.87$  ($q=1/\sqrt 2\approx 0.71$) within our model. 

Qualitatively, the phase diagram can be understood as a consequence of Andreev 
bound states occuring at the interface. A sign-change of the gap function along a quasiparticle trajectory 
gives rise to the formation of zero-energy Andreev bound states. Such sign changes occur as soon as
$ q > 0 $. Changing the phase difference $ \phi $ from zero to a finite value may move these bound states away from
zero energy, resulting in an energy gain accordingly. For larger values of $q$ the original spectral weight of the 
zero-energy Andreev bound states gets enhanced. Consequently, upon increasing $q$ a continous transition to 
a state of finite $\phi$ occurs at some critical value $ q_c(T) $. If  $ q $ is increased further, $ \phi $ eventually 
reaches  the upper limiting value $ \pi $.

The extent of the intermediate region in the phase diagram depends on the transparency
of the intrinsic interface. The results presented sofar have been derived under
the assumption of a high-transparency metallic junction. For comparison,
we also examined the Josephson current for the opposite limit of a 
low-transparency tunnel junction. Employing standard 
boundary conditions for the quasiclassical propagators 
\cite{Eschrig,Zaitsev,Shelankov}, we find 
\begin{equation}
\label{EQCurrentLow}
\hat{\bf j}(\phi) 
=D\sin\phi \cdot \hat{T}\sum_{\omega_n >0}^{\omega_c} \int_0^{\frac{\pi}{2}} d\theta 
\left\{
\begin{array}{r}    \frac{4}{\pi} \sin^2\theta \; c_\perp \\  \sin 2 \theta \; c_\parallel   \end{array}  
\right\} \hat{ \mathbf{n}},
\end{equation} 
where the following notation is used 
\begin{subequations}
\begin{eqnarray}
c_\perp &=& \frac{-\gamma_- \gamma_+}{(1+\gamma_- \gamma_+)(\gamma_- -\gamma_+)} \arctan \frac{\gamma_- -\gamma_+}{1+\gamma_- \gamma_+}\\
c_\parallel &=& \frac{-\gamma_- \gamma_+}{(1+\gamma^2_-)(1+\gamma^2_+)}.
\end{eqnarray}
\end{subequations}
These results are valid to first order in the transparency $D\ll1$, and, as in Eq. (\ref{EQCurrent}) for the 
metallic interface, the upper formula corresponds to the orientation $\hat{\mathbf{n}}\perp\hat{\mathbf{z}}$
and the lower one to $\hat{\mathbf{n}}\parallel\hat{\mathbf{z}}$. The main difference to the metallic 
case can be seen quite clearly: Since the current-phase relation in Eq. (\ref{EQCurrentLow}) is purely sinusoidal,
the stable phase of the junction must be either $0$ or $\pi$, depending on the sign
of the amplitude factor. In particular, the intermediate region has been shrunk to
a single boundary line in the phase diagram.
In Figs. \ref{Fig02} and \ref{Fig03}, these boundaries between $0$- and $\pi$-regions 
in the tunnel limit are sketched by the dashed lines for comparison.

In the following, we concentrate on one remarkable physical consequence of
the intermediate region where $ \phi \neq 0, \pi $. The degeneracy of the two phases 
$ \pm \phi $ gives rise to the possibility of  line defects on the interface
which carry fractional magnetic flux. They can exist at the interface only, and may generally
exhibit fractional flux quanta $\Phi$ according to
\begin{equation}
\label{EQfracflux}
\frac{\Phi}{\Phi_0}=n\pm\frac{\phi}{\pi} \qquad \qquad n \in \mathbb{Z},
\end{equation}
where $ \Phi_0 =hc/2e $ is the standard flux quantum. 
 As a consequence of this property, it is possible for 
a standard vortex to decay into two fractional ones on the interface, carrying the 
fractional flux $\phi/\pi\cdot \Phi_0$ and $(1-\phi/\pi)\cdot\Phi_0$, respectively. Both of
these line defects are strongly pinned to the interface. If there are many of these fractional vortices lined up along the interface, 
they can act as a severe impediment for flux flow. 
Similar theoretical considerations have been made for domain walls in
time reversal symmetry breaking superconductors  \cite{Volovik,SigristUeda,SigristAgterberg}.

In summary, we find that interfaces between twin domains in a noncentrosymmetric
superconductor such as CePt$_3$Si could possess unusual properties. They can
host low-energy Andreev bound states and, under certain conditions, give rise to a
time reversal symmetry violating phase, a characteristic phase of the interface only. 
In this situation, fractional vortices could exist on the interface and severly influence 
the flux creep.
Since the interface properties are different for different orientations, the flux creep
properties would likely depend on the vortex direction. Furthermore, the change of the phase $ \phi $ 
across such an interface can also modify special interference features of the Josephson
effect in a magnetic field, if the interface intersects the junction between a noncentrosymmetric 
and a conventional superconductor. The low-energy Andreev bound states may be directly accessible 
 by local tunneling probes such as scanning tunneling microscopes. 

We would like to thank D.F. Agterberg, N. Hayashi, Y. Kitaoka, H. Mukuda and Y. Onuki for 
stimulating discussions. This work was financially supported by the Swiss Nationalfonds and
the NCCR MaNEP, as well as the Center for Theoretical Studies of ETH Zurich.

\end{document}